\documentclass[aps,prb,preprint,groupedaddress,showpacs]{revtex4}

\begin{document}


\title{Perturbative results on localization for a driven two-level system}


\author{Marco Frasca}
\email[e-mail:]{marcofrasca@mclink.it}
\affiliation{Via Erasmo Gattamelata, 3 \\
             00176 Roma (Italy)}


\date{\today}

\begin{abstract}
Using perturbation theory
in the strong coupling regime, that is, the dual Dyson series, and renormalization group
techniques to re-sum secular terms, we obtain the perturbation series of the
two-level system driven by a sinusoidal field till second order.
The third order correction to the energy levels is obtained proving how this
correction does not modify at all the localization condition for a strong field
as arising from the zeros of the zero-th Bessel function of integer order. 
A comparison with weak coupling perturbation theory is done showing how
the latter is contained in the strong coupling expansion in the proper limits.
The strong coupling expansion we obtain proves to be accurate in the regime of
high-frequency driving field.
This computation gives an explicit analytical form to Floquet eigenstates and quasi-energies
for this problem,
for high-frequency driving fields,
supporting recent theoretical and experimental findings
for quantum devices expected to give a representation for qubits in quantum computation.
\end{abstract}


\maketitle

\section{Introduction}

Quantum computation\cite{qc1,qc2,qc3,qc4}, intended as information theory performed using
qubits, that is, two-level systems evolving by the unitary evolution as obtained by the
laws of quantum mechanics, demands an increasing control on such systems.

Recent experimental findings\cite{ex5,ex6,ex7,ex8,ex9} proved that qubits can be realized
by solid state devices, making the realization of a quantum
computer even more possible. Such solid state devices performs in a way or
another like two-level systems driven externally by some time-varying field.
This means that it is very important a clear theoretical understanding of them
to make their control easier.

Advances in this way have been realized by devising new approaches to the
solution of the Schr\"odinger equation in different regimes\cite{mt9,mt10,mt11,mt12,mt13,mt14}.
These methods are perturbative in their nature but permit to study a quantum
system in different physical regimes. So, they seems to be a proper framework to
make two-level systems theoretically manageable in any situation. These methods
give to the Floquet method, for periodical perturbations, a strong analytical
support.

The proper control of a qubit, realized by some solid state device, relies
on the possibility to obtain localization of a particle in each one of the two
available states. This effect is also known as \textit{coherent destruction
of tunneling} (CDT)\cite{hanggi1,hanggi2} as, tunneling between one state and the other
is destroyed by making unity the probability of staying of the particle in the 
initial state. So, this question has been recently the main motivation for
several works on two-level systems\cite{tl15,tl16,tl17,tl18}. Some of these papers have put forward
an interesting non perturbative result showing a set of curves for the crossing
of the quasi-energies of the Floquet solution for two-level systems with different
periodical external fields. As it stands, such a result, being not perturbative in
its very nature, appears rather difficult to recover with perturbation theories.

The aim of this paper is to show how perturbation theory can give definite results
for localization in different regimes, putting the study of two-level systems on
a sound ground from an analytical point of view. Besides, an expression is given
for localization, after that an effective time-independent Hamiltonian is
obtained using perturbation theory. Our results confirm till third order the
well-known result that the zeros of the zero-th order Bessel function are the
points where CDT occurs. Besides, our result recovers the small
perturbation theory computation in the proper limits. This is in agreement with
the recent numerical computations \cite{tl17,tl18}.
But, it is essential to emphasize that the main result of the paper, the
strong coupling expansion, is very accurate for high-frequency driving fields.

The paper is so structured. In Sec. \ref{sec2} we discuss localization on a
quite general ground, using Floquet theory. In Sec. \ref{sec3} we give a
presentation of the perturbation methods we use to obtain our results. In
Sec. \ref{sec4} we present the computation for the two-level system driven by
a sinusoidal field giving the effective time-independent Hamiltonian obtained
by a computation till second order. In Sec. \ref{sec5} we give the third order
correction to the effective Hamiltonian showing how the localization condition at
high fields is kept in agreement with all numerical results as also happens to
the small field condition. Then we discuss the conditions under which localization
happens comparing the perturbation series obtained
with the weak perturbation series showing the way one result contains the other. Finally,
in Sec. \ref{sec6} conclusions are given.

\section{Localization in driven two-level systems\label{sec2}}

The model we consider has the Hamiltonian
\begin{equation}
    H = \frac{\Delta}{2}\sigma_3+g\sigma_1 f(t) \label{eq:H}
\end{equation}
being $\Delta$ the separation between the two levels, $g$ the coupling,
$\sigma_1$ and $\sigma_3$ Pauli matrices and $f(t)=f(t+T)$ with $T$ the period
of the perturbation. The theory of CDT \cite{hanggi1,hanggi2} can be straightforwardly
applied. Given the unitary evolution operator $U(t)$, one can compute the probability
of the system of being in the initial state as
\begin{equation}
    P(t) = |\langle\psi(0)|U(t)|\psi(0)\rangle|^2
\end{equation}
with$|\psi(0)\rangle$ being one of the eigenstates of $\sigma_3$. One has CDT when
this probability is equal to unity. As the Hamiltonian is periodic in time, one
can apply the Floquet theory and take\cite{Hanggi3}
\begin{equation}
    \bar P=\frac{1}{T}\int_0^T P(t)dt
\end{equation}
to ascertain that we have true localization by having unity also in this case. The
consequence is that one can show, again from Floquet theory, that the conditions
for CDT arise from crossing of quasi-energies. This condition is necessary but not
sufficient.

Our aim in the following is to derive such conditions that apply to the 
quasi-energies for the sinusoidal case, that is, assuming $f(t)=\cos(\omega t)$
with $\omega = 2\pi/T$. We give an explicit analytical
expression for the Floquet eigenstates and quasi-energies by using perturbation
theory in the strong coupling regime with the method described in the next section.

\section{Dual Dyson series and resummation techniques \label{sec3}}

It is generally believed that perturbation theory, being plagued by secularities
, that is, unbounded terms in a perturbation series that increase as powers of time, $t,t^2,\cdots$,
should be avoided
to treat
a model
like the one we consider in this
paper. Indeed, the methods devised so far for removing such singular terms proved
to be generally
not very easy
to apply. Recently, a new approach, that can be called
dynamical renormalization group method \cite{rg1,rg2,rg3,rg4,rg5,rg6}, has given an algorithmic
way to remove secularities, making computation in perturbation theory
just tedious but very easy to accomplish. This method, coupled with the
dual Dyson series \cite{fra1,fra2,fra3,fra4,fra5}, gives a straightforward method
to compute higher order corrections to well known results \cite{hanggi2}. Besides,
we will get an explicit analytical expression for the Floquet modes and the
quasi-energies in the given approximation and we will prove that
the small coupling result is recovered. This is a property of the dual Dyson series.

The dynamical renormalization group method was firstly formulated by the Urbana
Group \cite{rg1,rg2} based on the observation that renormalization group methods
can be seen as a means of asymptotic analysis \cite{rg1}. They showed how their
approach can be proven equivalent to other methods that resum secularities,
such as the multiple time scale method \cite{kc}, for many cases. This method,
for its very nature, can be described by an example. So, let us consider the
well-known equation of the forced harmonic oscillator
\begin{equation}
   \ddot x(t) = -x(t)-\epsilon x(t)^3
\end{equation}
being $\epsilon$ the strength of the anharmonic term assumed to be small. A naive
perturbation expansion in $\epsilon$ till first order gives the well-known result
\begin{equation}
\label{eq:xn}
    x(t) = A_0\sin(t+\phi_0)+\epsilon\frac{3}{8}A_0(t-t_0)\cos(t+\phi_0)
	-\epsilon\frac{A_0}{16}\sin(3t+3\phi_0)+O(\epsilon^2) 
\end{equation}
being $A_0$ and $\phi_0$ constants depending on the initial conditions. We recognize
a secular term $t-t_0$ that makes this series useless as it breaks down for
$\epsilon(t-t_0)>1$. The situation can be improved if we interpret the time
$t_0$ as the logarithm of the ultraviolet cutoff in quantum field theory and
introduce $A$ and $\phi$ as the renormalized counterpart of $A_0$ and $\phi_0$
due to the fact that the nonlinearity may change these constants. Then, we
introduce another renormalization point by splitting $t-t_0$ in
$t-\tau + \tau - t_0$ and adsorb the terms containing $\tau-t_0$ into
$A$ and $\phi$. We introduce a multiplicative renormalization constant
$Z_1=1+\sum_{n=1}^\infty a_n \epsilon^n$ and an additive one  
$Z_2=\sum_{n=1}^\infty b_n \epsilon^n$ so that $A_0(t_0)=Z_1(t_0,\tau)A(\tau)$
and $\phi_0(t_0)=\phi(\tau)+Z_2(t_0,\tau)$ with the coefficients $a_n$ and $b_n$
to be computed order by order to remove the terms with $\tau-t_0$ as happens
in standard renormalization group \cite{gl1,gl2}. It is easily seen that, in our case,
a possible choice to first order is $a_1=0$ and $b_1=-\frac{3}{8}(\tau-t_0)$ 
removing the secular term and we are left with
\begin{equation}
\label{eq:x}
    x(t)=A\sin(t+\phi)+\epsilon\frac{3}{8}A(t-\tau)\cos(t+\phi)
	-\epsilon\frac{A}{16}\sin(3t+3\phi)+O(\epsilon^2).
\end{equation}
being now $A$ and $\phi$ function of $\tau$. But $\tau$ does not appear in the
original problem and so $x(t)$ must be independent on it. This gives us the condition 
\begin{equation}
    \left.\frac{\partial x}{\partial\tau}\right|_{\tau=t}=0
\end{equation}
that is the renormalization group equation that in our case gives the equations
\begin{eqnarray}
    \frac{\partial A}{\partial t}&=& O(\epsilon^2) \\ \nonumber
	\frac{\partial\phi}{\partial t} &=& \frac{3}{8}\epsilon A+O(\epsilon^2)
\end{eqnarray}
where the well-known shift in the frequency of the oscillator is recovered. Besides,
the secularity is completely removed by taking into eq.(\ref{eq:x}) the
condition $\tau=t$ with $A(t)$ and $\phi(t)$ given by the solutions of the
renormalization group equations. We have finally
\begin{equation}
    x(t)=A(0)\sin\left[\left(1+\frac{3}{8}\epsilon A(0)\right)t+\phi(0)\right]
	-\epsilon\frac{A(0)}{16}\sin\left[3\left(1+\frac{3}{8}\epsilon A(0)\right)t+3\phi(0)\right]+O(\epsilon^2).
\end{equation}
So, a straightforward application of renormalization group methods has permitted
to remove a secular term in the perturbation series. But there is a way to make
the computation simpler by noting that we have done nothing else than to compute
the envelope of eq.(\ref{eq:xn}) using known renormalization techniques.

Then,
the
method of renormalization group to resum secularities in a perturbation series that we 
present here is obtained by the mathematical theory of envelopes and is due to Kunihiro \cite{rg3,rg4,rg5}.
In order to describe it, let us consider the following equation
\begin{equation}
    \dot x(t)=f(x(t),t), \label{eq:xf}
\end{equation}
with $x(t)$ that can be also a vector. The initial condition is given by $x(t_0)=X(t_0)$.
At this stage we assume $X(t_0)$ not yet specified. We write the solution of
this equation as $x(t;t_0,X(t_0))$ that is exact. If we change $t_0$ to $t_0'$
we are able to determine $X(t_0)$ by assuming that the solution should not change
\begin{equation}
    x(t;t_0,X(t_0)) = x(t;t_0',X(t_0'))
\end{equation} 
that in the limit $t_0\rightarrow t_0'$ becomes
\begin{equation}
     \frac{dx}{dt_0}=\frac{\partial x}{\partial t_0}+
	 \frac{\partial x}{\partial X}\frac{\partial X}{\partial t_0}=0
\end{equation}
giving the evolution equation or flow equation of the initial value $X(t_0)$.
We see again the renormalization group equation proper to this approach.

Till now, all our equations are exact and no perturbation theory entered in any part
of our argument. But, except for a few cases, the solution $x(t;t_0,X(t_0))$ is
only known perturbatively and such a solution is generally valid only locally,
i.e. for $t\sim t_0$ and $t\sim t_0'$ and a more restrictive request should be
demanded to our renormalization group equation
\begin{equation}
     \left.\frac{dx}{dt_0}\right|_{t_0=t}=
	 \left.\frac{\partial x}{\partial t_0}\right|_{t_0=t}+
	 \left.\frac{\partial x}{\partial X}\frac{\partial X}{\partial t_0}\right|_{t_0=t}=0. \label{eq:rg}
\end{equation}
But this equation can be interpreted by the mathematical theory of envelopes \cite{rg3}.
Indeed, varying $t_0$ we have that $x(t;t_0,X(t_0))$ is a family of curves
with $t_0$ as a characterizing parameter. Then, eq.(\ref{eq:rg}) becomes
an equation to compute the envelope of such a family of curves. Such an envelope
is given by $x(t;t_0=t)=X(t)$, the initial condition. It can be proven that
$X(t)$ satisfies the equation (\ref{eq:xf}) in a global domain up to the order
$x(t;t_0)$ for $t\sim t_0$.

Now,
we
can
describe our approach in some steps to show how such an algorithmic perturbation 
method indeed works
\begin{enumerate}
\item Consider the following unitary transformation on the Hamiltonian (\ref{eq:H}) 
(here and in the following we set $\hbar = 1$) to remove the perturbation \cite{fra1,fra2}
\begin{equation}
     U_F(t) = \exp\left[-ig\sigma_1\int_0^t f(t')dt'\right]
\end{equation}
giving the transformed Hamiltonian
\begin{equation}
     H_F(t) = \frac{\Delta}{2}\sigma_3\exp\left[-i2g\sigma_1\int_0^t f(t')dt'\right];
\end{equation}
the dual Dyson series is computed by \cite{fra1,fra2,fra3,fra4}
\begin{equation}
     S_D(t,t_0) = {\cal T}\exp\left[-i\epsilon\int_{t_0}^tH_F(t')dt'\right]
\end{equation}
being as usual ${\cal T}$ the time ordering operator and an ordering parameter $\epsilon$
has been introduced that will be taken unity at the end of computation. It is fundamental for our
argument that the computation of this series is performed at a different starting point $t_0$.
\item Assume, at the start, that the time evolution operator has the form
\begin{equation}
    U(t,t_0) = U_F(t)S_D(t,t_0)U_R(t_0)
\end{equation}
where $U_R(t_0)$ is a ``renormalizable'' part of the unitary evolution.
\item At the given order one gets $S_D(t,t_0)$ as
\begin{equation}
    S_D(t,t_0) = I - i\epsilon f_1(t,t_0) -\epsilon^2 f_2(t,t_0)+\ldots
\end{equation}
and, at this stage, if some oscillating functions in $t_0$ appear like $e^{-i\omega t_0}$
then introduce the phase $\phi(t_0) = -t_0$ as a ``renormalizable'' parameter
rewriting it as $e^{i\omega\phi(t_0)}$\cite{rg6}.
We make this choice assuming that any initial phase of the system can be changed by the
dynamics. The minus sign is fixed arbitrarily.
The secularities must be left untouched.
\item Eliminate the dependence on $t_0$ by requiring\cite{rg3,rg4}
\begin{equation}
     \left.\frac{dU(t,t_0)}{dt_0}\right|_{t_0=t} = 0 \label{eq:rgcond}
\end{equation}
and one obtains the renormalization group equation
\begin{eqnarray}
    \frac{dU_R(t)}{dt} &=& \epsilon g_1 U_R(t) + \epsilon^2 g_2 U_R(t) + O(\epsilon^3) \\ \nonumber
	\frac{d\phi(t)}{dt} &=& \epsilon \phi_1 \phi(t) + \epsilon^2 \phi_2 \phi(t) + O(\epsilon^3) 
\end{eqnarray}
where, at some stage, to obtain such equations at the second order, we have to use
their expressions at the first order
and,
to compute their form at order $n$-th, one
has to use these equations at the order $(n-1)$-th, into the condition (\ref{eq:rgcond}).
This is a step toward the computation of the envelope of the perturbation series \cite{rg3,rg4}.
\item Finally, the renormalization equations should be solved and substituted into the
equation
\begin{equation}
     \left.U(t,t_0)\right|_{t_0=t} \label{eq:u}
\end{equation}
giving the solution, i.e. the envelope, we were looking for without secularities at the order
we made the computation.
\end{enumerate}

We will give an explicit application of this procedure in the next section
choosing $f(t)=\cos(\omega t)$, that is, a sinusoidal driving.

\section{Structure of the perturbation series to second order \label{sec4}}

The unitary transformation $U_F(t)$ for a sinusoidal driving
field
takes the form
\begin{equation}
    U_F(t) = \exp\left[-i\frac{g}{\omega}\sin(\omega t)\sigma_1\right]
\end{equation}
and the transformed Hamiltonian takes the form
\begin{equation}
   H_F(t) = \frac{\Delta}{2}\sigma_3\exp\left[-iz\sin(\omega t)\sigma_1\right].
\end{equation}
having put $z=\frac{2g}{\omega}$.
This Hamiltonian gives the term $S_D(t,t_0)$ till first order as
\begin{eqnarray}
     S_D(t,t_0) &=& I - i\frac{\Delta}{2}\sigma_3 J_0(z)(t-t_0) \\ \nonumber
	 & &+\frac{\Delta}{2}\sigma_3\sigma_1
	 \sum_{n\neq 0}J_n(z)\frac{e^{-in\omega t\sigma_1}-e^{-in\omega t_0\sigma_1}}{n\omega}
	 +\ldots 
\end{eqnarray}
where $J_n(z)$ are the Bessel functions of $n$-th order being $n$ an integer.
We see immediately that a secular term, proportional to $t-t_0$, plagues
our computation. To remove it we rewrite the above expression as
\begin{eqnarray}
     S_D(t,t_0) &=& I - i\frac{\Delta}{2}\sigma_3 J_0(z)(t-t_0) \\ \nonumber
	 & &+\frac{\Delta}{2}\sigma_3\sigma_1
	 \sum_{n\neq 0}J_n(z)\frac{e^{-in\omega t\sigma_1}-e^{in\omega\phi(t_0)\sigma_1}}{n\omega}
	 +\ldots 
\end{eqnarray}
introducing into the last oscillating term the renormalizable parameter $\phi(t_0)$.
So, one has
\begin{eqnarray}
    \frac{dU(t,t_0)}{dt_0} &=& U_F(t)  \label{eq:fo}
	\left[i\frac{\Delta}{2}\sigma_3 J_0(z) -
	 \frac{\Delta}{2}\sigma_3
	 \sum_{n\neq 0}J_n(z)e^{in\omega\phi(t_0)\sigma_1}\frac{d\phi(t_0)}{dt_0}
	\right]U_R(t_0) \\ \nonumber
	 &+&U_F(t)\left[I - i\frac{\Delta}{2}\sigma_3 J_0(z)(t-t_0)
	 + \frac{\Delta}{2}\sigma_3\sigma_1
	 \sum_{n\neq 0}J_n(z)\frac{e^{-in\omega t\sigma_1}-e^{in\omega\phi(t_0)\sigma_1}}{n\omega} + \ldots\right]
	 \frac{dU_R(t_0)}{dt_0} 
\end{eqnarray}
and imposing the condition (\ref{eq:rgcond}) we get the renormalization group equations
\begin{eqnarray}
    \frac{dU_R(t)}{dt} &=& -i\frac{\Delta}{2}\sigma_3 J_0(z) U_R(t) + 
	O\left[\left(\frac{\Delta}{\omega}\right)^2\right] \\ \nonumber
	\frac{d\phi(t)}{dt} &=& O\left[\left(\frac{\Delta}{\omega}\right)^2\right]
\end{eqnarray}
where use has been made of the fact that these equations are at least first order, transforming
some terms in eq.(\ref{eq:fo}) into second order, making them negligible. We see that we have recovered a well known
result of the series for high frequency\cite{hanggi2} with its corrections to first order. This
gives the final result, using eq.(\ref{eq:u}),
\begin{equation}
    U(t,0) = U_F(t)\left[I
	 + \frac{\Delta}{2}\sigma_3\sigma_1
	 \sum_{n\neq 0}J_n(z)\frac{e^{-in\omega t\sigma_1}-1}{n\omega} + \ldots\right]
	 e^{-i\frac{\Delta}{2}\sigma_3 J_0(z)t}.
\end{equation}
This result is well-known in literature\cite{fra3,fra5,mt10,mt11}. It is also been
proven that this is the expression of the Floquet unitary evolution at this order with
the given approximations\cite{mt10,mt11}. We see that we have a periodic part and
a part that originates the quasi-energies. 

So, what is really interesting is to get higher
order corrections. Particularly, we see that the quasi-energies are given by
$\pm\frac{\Delta}{2}J_0(z)$, a well-known fact, so that, CDT occurs, in this
approximation, at the zeros of the Bessel function $J_0(z)$\cite{hanggi2}. We now
prove that higher order corrections preserve such a result.

Applying the above procedure till second order it is straightforward to obtain
\begin{eqnarray}
     U(t,0) &=& U_F(t)\left[I + \frac{\Delta}{2}\sigma_3\sigma_1
	 \sum_{n\neq 0}J_n(z)\frac{e^{-in\omega t\sigma_1}-1}{n\omega}\right. \\ \nonumber
	 &-&i\frac{\Delta^2}{2}\sigma_1 J_0(z)\sum_{n\neq 0}J_n(z)\frac{\sin(\omega t)}{n^2\omega^2}
	 -\frac{\Delta^2}{4}\sum_{n\neq 0}J_n^2(z)\frac{e^{in\omega t}-1}{n^2\omega^2} \\ \nonumber
	 &-&\left.\frac{\Delta^2}{4}\sum_{n_1\neq 0,n_2\neq 0,n_1\neq n_2}
	 J_{n_1}(z)J_{n_2}(z)\left(\frac{e^{i(n_1-n_2)\omega t\sigma_1}-1}{n_2(n_1-n_2)\omega^2}
	 -\frac{e^{in_1\omega t\sigma_1}-1}{n_1 n_2\omega^2}\right) + \ldots  
	 \right]\times \\ \nonumber
	 & &e^{-i\frac{\Delta}{2}\sigma_3 J_0(z)t+i\frac{\Delta^2}{2\omega}F(z)\sigma_1 J_0(z)t}
\end{eqnarray}
where we have introduced the function $F(z)=\sum_{n\neq 0}\frac{J_n(z)}{n}$ into the last exponential.
We recognize a product of a periodic unitary operator and a term originating quasi-energies. So, 
it is interesting to see that now we have an effective Hamiltonian giving the quasi-energies with
a correction term proportional to $\sigma_1$. But, the most important point is that this
correction is again proportional to $J_0(z)$ and so, the zeros of this Bessel function
gives CDT also at second order. The effective Hamiltonian can be written as
\begin{equation}
    H_{eff} = \frac{\Delta J_0(z)}{2}\sigma_3-\frac{\Delta^2 J_0(z)}{2\omega}F(z)\sigma_1 \label{eq:heff}
\end{equation}
that can be easily diagonalized but we do not pursue this matter here. Indeed, our aim
in the next section will be to give the third order correction to this Hamiltonian proving
that the high-frequency CDT is again determined by the zeros of $J_0(z)$.

\section{Third-order correction to the quasi-energies and localization \label{sec5}}

The algebra is rather tedious but straightforward and the application of our method
gives the third order correction to the effective Hamiltonian
\begin{equation}
    H_{eff}^{(3)}= -\frac{\Delta^3 J_0(z)}{4\omega^2}\sum_{n\neq 0}\frac{J_n^2(z)}{n^2}\sigma_3. 
\end{equation}
This is a correction to the $\sigma_3$ term of the effective Hamiltonian so, we
can conjecture that even order corrections go to the $\sigma_1$ term and the
odd order corrections go to the $\sigma_3$ term. We see again that this term is
proportional to $J_0(z)$ confirming the exactness in the high-frequency limit
of the occurring of the CDT at the zeros of such a Bessel function. Such 
correcting terms into the effective Hamiltonian can be recognized as a.c. Stark
shifts and Bloch-Siegert shifts and are relevant, eventually, at the resonance\cite{rg6,fra6}
where Rabi flopping also in this regime is expected with the renormalized
levels $\pm\frac{\Delta}{2} J_0(z)$. 

Anyhow, a further check can be obtained with the small coupling perturbation theory. Our
method can be applied again but now we use the interaction picture. So, if we introduce
the unitary transformation $U_I(t)=e^{-i\frac{\Delta}{2}\sigma_3 t}$, we get the
transformed Hamiltonian
\begin{equation}
    H_I = g\cos(\omega t)\sigma_1 e^{-i\Delta\sigma_3 t}
\end{equation}
and we can built the Dyson series, out of resonance $(\Delta\neq\omega)$, to second order as
\begin{eqnarray}
    U(t,0)&=&U_I(t)\left[I-\frac{g}{2}\sigma_1\sigma_3
	\left(\frac{e^{i(\omega-\Delta)\sigma_3t}-1}{\omega-\Delta}-
	\frac{e^{-i(\omega+\Delta)\sigma_3t}-1}{\omega+\Delta}\right)\right. \\ \nonumber
	&+&\frac{g^2}{4}\left(
	\frac{\cos(2\omega t)+i\sigma_3\frac{\Delta}{\omega}\sin(2\omega t)}{\omega^2-\Delta^2}
	-\frac{e^{i(\omega+\Delta)\sigma_3t}-1}{\omega^2-\Delta^2} \right.\\ \nonumber
	&+&\left.\left.\frac{e^{i(\omega-\Delta)\sigma_3t}}{(\omega-\Delta)^2}
	-\frac{e^{-i(\omega-\Delta)\sigma_3t}}{\omega^2-\Delta^2}
	+\frac{e^{i(\omega+\Delta)\sigma_3t}}{(\omega+\Delta)^2}
	-2\frac{\omega^2+\Delta^2}{(\omega^2-\Delta^2)^2}
	\right)+\ldots
	\right]e^{i\frac{g^2}{2}\frac{\Delta}{\omega^2-\Delta^2}\sigma_3t}.
\end{eqnarray}
A rather interesting result is that this series holds for any ratio $\frac{\Delta}{\omega}$
differently from the dual Dyson series discussed above that holds for
$\frac{\Delta}{\omega}\ll 1$. From the
effective Hamiltonian (\ref{eq:heff}) and the third order correction we see that
the coefficient of the $\sigma_3$ part is given by
\begin{equation}
    S_3 = \frac{\Delta J_0(z)}{2}-\frac{\Delta^3 J_0(z)}{4\omega^2}\sum_{n\neq 0}\frac{J_n^2(z)}{n^2}
\end{equation}
that in the small coupling approximation, being $J_0(z)\rightarrow 1-\frac{z^2}{4}$, and considering just
the first term of the series giving $\frac{z^2}{2}$, we obtain
\begin{equation}
    S_3\approx\frac{\Delta}{2}(1-\frac{z^2}{4})-\frac{\Delta^3}{4\omega^2}\frac{z^2}{2} + O(z^4). 
\end{equation}
From the small coupling expansion we recognize, after a simple
rearranging of the terms, the $\sigma_3$ term
\begin{equation}
    S'_3=\frac{\Delta}{2}-\frac{g^2}{2}\frac{\Delta}{\omega^2-\Delta^2} \label{eq:S'3}
\end{equation}
that, in the expected limit $\frac{\Delta}{\omega}\ll 1$, is the same as $S_3$. 
Similarly, one can verify the same result at any order of the two perturbation series. This
gives an {\it a posteriori} verification of our computations. Besides, it confirms
our point that the dual Dyson series is a high-frequency series while the Dyson
series holds for both the limits of high and low frequencies, but the result of the
Dyson series in the high-frequency limit is contained in the dual Dyson series when
the limit of small coupling $z\rightarrow 0$ is also taken. Indeed, the Dyson series
and the dual Dyson series can be taken to coincide in the limits of small $z$ and
$\frac{\Delta}{\omega}\ll 1$, giving an interesting relationship between the 
transformed Hamiltonians we use to obtain the perturbation series.

We are now in a position to discuss the CDT in the limit of small coupling $z\rightarrow 0$. 
For our aim, we consider the analytical expression of the quasi-energies obtained in Ref.\cite{zhao}.
This gives
\begin{equation}
   \epsilon_\pm = \pm\frac{\omega}{2\pi}\arccos\left[
   \cos\left(\pi\frac{\Delta}{\omega}\right) +
   \pi\frac{z^2}{4}\frac{\Delta}{\omega}
   \frac{\sin\left(\pi\frac{\Delta}{\omega}\right)}{1-\frac{\Delta^2}{\omega^2}}
   \right], \text{mod}(\omega) \label{eq:zhao}
\end{equation}
that yields, in the limit $z\ll 1$, eq.(\ref{eq:S'3}) proving again the correctness of our
computations. The important point is that eq.(\ref{eq:zhao}) contain the
result for CDT that the quasi-energies crosses for $\frac{\Delta}{\omega}=2n$ being $n$
an integer and this cannot happen for the perturbative result unless we are able
to resum the series. So, we can draw the conclusion that perturbation theory by Dyson series can prove
to be very effective for the study of CDT for strong fields and high-frequency regime.

\section{Discussion and conclusions\label{sec6}}

From the von Neumann-Wigner theorem \cite{vNW} one has to expect that the crossings
of the energy levels form a one-dimensional manifold. This result is central to
fully understand CDT for the two-level system we considered. Recent works on this
question derived such a result numerically \cite{tl17,tl18}. In this paper we have
shown that, unless some smart resummation technique is applied to a perturbation
series, perturbation theory is helpful just to study the behavior of the
model in different regimes, recovering CDT under certain conditions.

The reason for this conclusion lies on the fact that the content of the von Neumann-Wigner
theorem is non-perturbative. So, if we are able to resum a perturbation series to
recover a non-perturbative result, there may be a possibility to give an analytical proof
of the numerical results obtained so far.

Notwithstanding such a conclusion, we have proved the power of the Dyson series
in the study of two-level systems in regimes where other means may prove unsuccessful,
giving explicitly the form of the Floquet modes till second order and quasi-energies
till third order, for a strong field and high frequency. At this stage, it appears
mandatory to introduce resummation techniques in perturbation theory to complete
the algorithmic approach we discussed in this paper.

Different regimes of parameter space have been analyzed by our perturbation
technique. Notably, we have obtained perturbation series for the high-frequency
regime $\Delta\ll\omega$ in the strong coupling approximation $g\gg\Delta$ by
dual Dyson series. The Dyson series has given the small coupling expansion
$g\ll\Delta$ that holds for any ratio $\frac{\Delta}{\omega}$ but when this ratio
is taken to be small, the dual Dyson series recovers the Dyson series for
$g\ll\Delta$ as we have shown. These results appear very promising for the
study of quantum systems in different regimes by a general way to approach
computations with perturbation theory.
\begin{acknowledgments}
I would like to thank Charles Creffield for suggesting me such
an interesting question.
\end{acknowledgments}


\end{document}